\documentstyle[aps,prl,floats,psfig]{revtex}
\tighten
\draft
\begin{document}
\twocolumn[\hsize\textwidth\columnwidth\hsize\csname
@twocolumnfalse\endcsname



\title{Is there $np$ pairing in odd-odd N=Z nuclei?}

\author{A.O.~Macchiavelli, P.~Fallon,
R.M.Clark,  M.Cromaz, M.A.Deleplanque, R.M.Diamond,
G.J.Lane, I.Y.Lee, F.S.Stephens, C.E.Svensson, K.Vetter, and D.Ward.}
\address{ Nuclear Science Division, Lawrence Berkeley National Laboratory,
Berkeley, CA 94720\\}
\date{\today}
\maketitle

\begin{abstract}
The binding energies of even-even and odd-odd $N=Z$ nuclei are compared.
After correcting for the symmetry energy we find that the lowest
$T=1$ state in odd-odd $N=Z$ nuclei is as bound as the ground state in the 
neighboring even-even nucleus, thus providing evidence for isovector $np$
pairing.
However, $T=0$ states in odd-odd $N=Z$ nuclei are several MeV less bound than
the even-even ground states.
We associate this difference with a pair gap and conclude that there is no
evidence for isoscalar correlated pairs in $N=Z$ nuclei.

\end{abstract}
\vspace{0.5cm}]


\narrowtext

Soon after the interpretation of superconductivity in terms of a condensate of
strongly correlated electron pairs (Cooper pairs) by Bardeen, Cooper and
Schrieffer (BCS)\cite{BCS} a similar pairing mechanism was invoked
for the  nucleus\cite{BMP} to explain, for example, the energy gap in 
even-even nuclei and the magnitudes of moments of inertia.
For almost all known nuclei, i.e.~those with $N>Z$, the ``superfluid'' state
consists of neutron ($nn$) and/or proton
($pp$) pairs  coupled to angular momentum zero and isospin T=1.
However, for nuclei with  $N=Z$ the near degeneracy of the proton and 
neutron Fermi
surfaces (protons and neutrons occupy the same orbitals) leads to a
second class of Cooper pairs consisting of a neutron and a proton ($np$).
The $np$ pair can couple to angular momentum zero and isospin $T=1$ (isovector), 
or, since they
are no longer restricted by the Pauli exclusion principle, they can couple to $T=0$ 
(isoscalar) and the angular momentum is $J=1$ or $J=J_{max}$ \cite{Schiffer},
 but most commonly the maximum value \cite{Zeldes}. 
Charge independence of the nuclear force implies that for $N=Z$ nuclei, 
$T=1$ $np$ pairing should exist on an equal footing with  $T=1$
$nn$ and $pp$  pairing.
Whether there also exists  strongly correlated $T=0$ $np$ pairs, has
remained an open question.
Early theoretical works\cite{Goodman} discussed the competition between 
$T=0$  and $T=1$ pairing within the BCS framework.
Recent works have focussed on the solutions of schematic (or algebraic) \cite{Engel,Isacker}
and realistic shell models \cite{Poves}, as well as on 
the properties of heavier $N=Z$ nuclei\cite{goodman98},
and the effects of rotation\cite{satula,stefan}.

To date, there exists a wealth of experimental evidence in support of 
the existence of $nn$ and $pp$ pairs, but little or no  evidence for
$np$ pairing mainly because of the experimental difficulties in studying $N=Z$
nuclei. Nevertheless, 
following recent advances in the experimental techniques and considering the new 
possibilities that will become available with radioactive beams, 
there has been a revival of nuclear structure studies along the $N=Z$ line. 
In this letter we present 
an analysis\footnote{In preparing this manuscript, a preprint (P.Vogel, Los Alamos
preprint, nucl-th/980515) came to our attention describing a very similar analysis to that 
presented here and with similar conclusions.}
of experimental binding energies ($BE$)  of nuclei along the
$N=Z$ line and the relative excitation energies of 
the lowest $T=0$ and $T=1$ states in 
self-conjugate ($N=Z$, $T_z=0$) odd-odd nuclei.
We have found evidence for the existence of strong $T=1$ $np$ pairing in 
$N=Z$ nuclei, but find no such evidence for  $T=0$ $np$ pairing.

Let us start by recalling that  pairing effects can be isolated by studying
differences in binding energies\cite{bm}. Particularly, the difference
\begin{eqnarray}
BE_{even-even}-BE_{odd-odd} \approx \Delta_p + \Delta_n \approx 2\Delta
\end{eqnarray}
is used as a measure of the pair gap, $\Delta$, for both protons and 
neutrons\footnote{There is usually a 
correction term due to the residual $np$ interaction. 
This term is of order $20~\mbox{MeV}/A$ and 
we will not consider it here.}.

Implicit in Eq. (1) is the assumption that the ground
states  have the same
isospin,  which is the case for nuclei with $N \neq Z$ 
since they are {\it maximally aligned}
in isospace, i.e. $T=T_z={1 \over 2}(N-Z)$\cite{bm}. 
Equation (1) is also true when comparing  $T=0$ states in even-even and
odd-odd $N=Z$ nuclei, and the difference
in binding energy,  given by
\begin{eqnarray}
BE_{ee}(N,Z)-{(BE_{oo}(N-1,Z-1)+BE_{oo}(N+1,Z+1)) \over 2},
\end{eqnarray}
is shown in Fig.~\ref{t0}, where the binding energies are from Ref.~\cite{wapstra}.
For comparison, the same quantities are given for nuclei with $N=Z+4$.
Taking the average in Eq.~(2) removes the smooth variations due to 
volume, surface, and Coulomb energies, and any
remaining differences are then attributed to shell or pairing effects. 
The extra binding of the even-even systems is clearly seen in 
Fig.~\ref{t0}  and it follows
the known  $1/A^{1/2}$ dependence\cite{bm}. 

\begin{figure}[htbp]
\centerline{\psfig{figure=paperfig1.epsi,height=6cm,width=8cm,angle=90}}
\caption{The difference in binding energy  between even-even and odd-odd nuclei.
Squares correspond to $T$=0 states in $N=Z$ nuclei ($T$=0 is the ground state
for $N=Z$ even-even nuclei and for  $N=Z$ odd-odd nuclei with $A<40$).
Circles correspond to the $T$=2 ground states in nuclei with N=Z+4.
In both cases this difference is interpreted as a measure of the pair gap, $\Delta$.
The solid line shows the conventionally adopted smooth dependence of $\Delta$ on A
($\Delta \sim 12\mbox{\mbox{MeV}}/A^{1/2}$).}
\label{t0}
\end{figure}

This result shows that the $T=0$ states in odd-odd $N=Z$ nuclei
behave  like those in any other odd-odd nucleus.
Assuming that the binding energy differences reflect differences 
in pairing energy then the extra $n$ and $p$ block the pairing to the same 
degree as  any ``standard'' 2-quasiparticle state.
Note,  if the ground states of $N=Z$ even-even nuclei  
contained $T=0$ correlated pairs, the addition of a $T=0$ $np$ pair would not 
give a gap, and the average binding energy of the two odd-odd 
nuclei  would be the same as the even-even neighbor.
This suggests that correlated $T=0$ pairs do not contribute significantly 
to the pairing energy in $N=Z$ nuclei.

Is it  possible that only $T=1$ pairing is important for these $N=Z$ nuclei? 
If $np$ $T=1$ pairs form a correlated state, the lowest $T=1$
state in self-conjugate odd-odd nuclei should be as bound as that of
the neighboring even-even ground state.
An analysis similar to that used for $T=0$ states should provide the
answer. However, in applying Eq.~(1) or (2) to determine the binding energy 
difference we need to include a symmetry energy term because of the different isospins 
(i.e.~$T=1$ in odd-odd $N=Z$ nuclei and $T=0$ in  neighboring even-even $N=Z$ nuclei).
A discussion on the symmetry term is given in refs.~\cite{bm,Blatt}. 
To extract the symmetry energy ($E_{sym} = -BE_{sym}$) the 
experimental binding energies
of several nuclei in the range $A=10-64$ were plotted, as shown in Fig.~\ref{sym}, after
subtracting volume, surface, and Coulomb terms. 
(The surface, Coulomb, and symmetry terms have the opposite sign to the volume term.)
They are plotted as a function of $T(T+x)$, for three cases: 
1) $x=4$, corresponding to the SU(4) Wigner supermultiplet expression\cite{Wig}, 
2) $x=1$, i.e.~$T(T+1)$,
and 3) $x=0$, giving a $T^2$ approximation. 
While any of these choices can  be used, the $T(T+1)$  
expression provides a better account of the 
experimental data, as discussed in Ref.~\cite{JAN}.
In our analysis we use a symmetry energy given by 
$E_{sym} = { 75\mbox{\tiny MeV} \over A} T(T+1)$ which
 represents an average  neglecting the effects of shell structure
and pairing.
\begin{figure}
\centerline{\psfig{figure=paperfig2.epsi,height=6cm,width=8cm,angle=90}}
\caption{The symmetry energy, $E_{sym} = -BE_{sym} = E_{exp}-E_{vol}-E_{surface} - E_{Coulomb}$, for nuclei in the range $A=10-64$ as a function of $T(T+x)$, as discussed in the text;
where, $E_{vol} = - 15A~\mbox{MeV}$, 
$E_{surface} = 17A^{2/3}~\mbox{MeV}$, and 
$E_{Coulomb} = 0.7 {Z^2 \over A^{1/3}}$ MeV.
Lines are to guide the eye.}
\label{sym}
\end{figure}
The binding energy difference for $T=1$ states in odd-odd $N=Z$ nuclei 
compared with $T=0$ ground states in
neighboring even-even $N=Z$ nuclei is presented in Fig.~\ref{t1} (squares). 
\begin{figure}
\centerline{\psfig{figure=paperfig3.epsi,height=6cm,width=8cm,angle=90}}
\caption{The difference in binding energy 
between even-even and odd-odd $N=Z$ nuclei.
The  $T$=1 states in odd-odd $N=Z$ nuclei (squares) are compared with the 
$T$=0 ground states
in neighboring even-even nuclei. The upper solid line represents the energy 
difference due to the difference in isospins ($T=1$ and $T=0$).
The dotted line was obtained from an average energy of $T=1$ isobaric analog states (see text).
The isospin correction term is seen to account for the observed binding energy 
difference.
The difference in binding energy is also shown for  odd nuclei, $Z=N+1$ (stars),
compared with the
even-even neighbor and the corresponding symmetry energy is given by the 
lower solid curve.
}
\label{t1}
\end{figure}
If the only difference between the even-even ground state 
and the odd-odd $T=1$ state were the symmetry term, then the difference in binding
energy is given by the upper solid line.
That is, the symmetry energy of the $T=1$ state 
(${75\mbox{\tiny MeV} \over A} T(T+1) = {150\mbox{\tiny MeV} \over A}$)  
subtracted from  the binding energy of the even-even nucleus  
provides the correct reference to which the odd-odd 
$T=1$ states should be compared. 
It is also possible to use the even-even $T=1$ ($T_z = -1, 1$) isobaric 
analog states  as a reference, rather than the global expression 
${75\mbox{\tiny MeV} \over A} T(T+1)$.
After correcting for the Coulomb energy, the binding energies of the isospin triplet 
are  very similar, often within a few hundred keV. 
The average binding energies of the  even-even $T=1$ ($T_z = -1, 1$)
isobaric analog states, relative to the even-even $T=0$ ground state, 
are also shown in Fig.~\ref{t1} (dotted line). 
These values are extremely close to those of the corresponding $T=1$, $T_z = 0$
state in the odd-odd nucleus.
Since, (i) the binding energy difference between the $T=1$, $T_z = 0$, (odd-odd)
and $T=0$, $T_z = 0$ (even-even) states is described by the symmetry energy term only,
and (ii)  the $T=1$ ($T_z = -1, 1$) state is the ground state of the even-even 
isobaric analog, then the binding energy difference ($BE_{ee}(T=0) - BE_{oo}(T=1)$) 
cannot be associated with a difference in pairing. 
Rather, it is due to the difference in isospin for which the smooth 
overall behavior is given by the symmetry energy.

These results indicate that the lowest $T=1$ state in a self conjugate odd-odd
nucleus is as bound as the neighboring even-even $N=Z$ ground state
(after correcting for the symmetry energy). 
In other words, there is no difference in pairing, and
just as the addition of an $nn$ or $pp$ pair to an even-even nucleus
does not block pair correlations, neither does the addition of an $np$ $T=1$ 
pair in $N=Z$ nuclei.
However, as expected, adding a single $n$ or $p$ to the even-even core 
does reduce the pair energy and results in a binding energy difference 
in excess of the symmetry energy, 
as seen by the fact that the data points (stars in Fig.~\ref{t1})
for an odd nucleus ($N=Z+1$) lie 
higher than the  symmetry energy expected for a T=1/2 nucleus
(lower solid curve in Fig.~\ref{t1}).
In view of the charge-independence of the nuclear force these results 
may not be too surprising; nevertheless they provide a strong argument in 
favor of the existence of full (i.e.~$nn$, $pp$, and $np$) 
isovector pairing correlations in $N=Z$ nuclei.

Finally, we consider the relative energies of the $T=0$ and $T=1$ states in 
odd-odd $N=Z$ nuclei.
If there were no $np$ pairing of any type ($T=0$ or $T=1$) the
$T=1$ state should lie above the $T=0$ state at an excitation energy given by the
symmetry term.
However, the analysis of the experimental data presented above shows strong 
evidence for the existence of $T=1$ $np$ pair correlations, and at 
the same time no evidence for $T=0$ correlated pairs.
The $T=1$ states should then lie at a lower energy than that  given by
the symmetry term, and if the $T=1$ pairing energy were sufficiently large, the 
$T=1$ state may lie lower than the $T=0$ state.
The experimental energy differences are shown in  Fig.~\ref{gap} along with
the expected contribution from the symmetry energy.
\begin{figure}
\centerline{\psfig{figure=paperfig4.epsi,height=6cm,width=8cm,angle=90}}
\caption{The difference in level energies between $T$=1 and $T$=0 states in odd-odd 
$N=Z$ nuclei.
For $A < 40$ these nuclei have $T=0$ ground states, except $^{34}$Cl,
 above this mass they have $T=$1 ground states, except $^{58}$Cu.
The solid line represents the isospin correction term 
$\Delta E_T = 150\mbox{MeV}/A$ and corresponds to the expected
energy difference between $T=1$ and $T=0$ states if the only difference between 
these states were due to the isospin correction term.
Insert: Squares denote the effective $\Delta_{np}$ gap derived from the relative 
excitation energies of the $T=0$ and $T=1$ states in odd-odd $N=Z$ nuclei 
after correcting for the isospin difference as illustrated in the main figure.
The solid line shows the result of a simple BCS calculation (see text for details).
}
\label{gap}
\end{figure}
The energy separation between the states of different isospin is clearly
less than that predicted by the symmetry term.
This is consistent with the pairing arguments presented above, and suggests that
whether the $T=0$ or $T=1$ state is lower depends
largely on the relative magnitudes of the symmetry and pairing energies.
We further note that while the near cancellation  of the
symmetry and pairing terms (for $T=1$ compared with $T=0$)
appears to be accidental we can not rule out, at this time,
a deeper physical origin.

Assuming the reduced separation is only due to the effects of pairing then,
in the language of the BCS model and taking the symmetry term  into 
account, the $T=0$ state in the odd-odd $N=Z$ nucleus 
can be interpreted as a 2-quasiparticle excitation (``broken-pair'' with seniority 2) 
relative to the $T=1$ correlated pair state. 
In complete analogy with Eq. (1) we have
\begin{eqnarray}
(BE_{T=1}-BE_{sym})-BE_{T=0} \approx 2\Delta_{np},
\end{eqnarray}
or, in terms of excitation energies,
\begin{eqnarray}
E_{sym}-(E_{T=1} - E_{T=0}) \approx 2\Delta_{np}.
\end{eqnarray}
(Note, this is the difference between the lowest $T=1$ and $T=0$ state in the same
N=Z odd-odd nucleus.)
The effective gap ($\Delta_{np}$), thus extracted, 
is presented in the insert to Fig.~\ref{gap}, 
where for comparison the result of a BCS calculation
that includes $nn$, $pp$, and $np$ $T=1$ pairs is also shown.
In this calculation we adopted standard 
single-particle levels from a spherical Nilsson potential  
and a pairing strength of $ 20 \mbox{MeV}/A$.
This figure illustrates that the magnitude of $2\Delta_{np}$ extracted from
experiment using Eq.~(4) compares
favorably with that obtained from the spectrum of single-particle levels.
While the gap (difference in binding energy) is not necessarily related only to a 
pairing  interaction \cite{Witek}, the agreement  is remarkable.

Due to the presence of shell gaps the simple BCS model gives a 
characteristic oscillation in $\Delta$. In this calculation, 
the single-particle levels were 
truncated at $N=Z=50$, which led to an artificial quenching of $\Delta$ at A=100. 
The reversal of the  favored isospin from $T=1$ to $T=0$ at $^{58}$Cu coincides with
it being one $np$ pair above the $N=Z=28$ and, within the pairing 
interpretation given here,
occurs because the shell gap reduces the magnitude of the  $T=1$ pair gap.
For heavier nuclei, $A>60$, the $T=1$ state is favored and we would expect 
that this is likely to remain the case until the $N=Z=50$ shell gap is 
reached, where for  $^{98}$In (N=Z=49) the ground state may well revert to 
$T=0$ once more.
The competition between pairing and symmetry energy was also
discussed in Ref.~\cite{Zeldes}. In this work, semi-empirical fits to the binding 
energies  suggested that for odd-odd $N=Z$ nuclei
beyond the $1f_{7/2}$ shell, pairing correlations will result in $T=1$
ground states.

In conclusion, we have argued that binding energy differences 
indicate that the lowest $T=1$ states in 
odd-odd $N=Z$ nuclei are as bound as their even-even neighbors, 
which provides strong evidence for the presence of isovector $np$ pairing.
There is, however, no similar evidence to support the existence of  $np$ 
isoscalar pair correlations. 
The intriguing switch from  $T=0$ to  $T=1$ ground states
in odd-odd $N=Z$ nuclei arises from a subtle competition between the 
symmetry energy and isovector pairing.
For $A > 40$, $T=1$ pairing wins over the symmetry energy and the $J=0^+$ 
state becomes
the ground state, except, possibly, near closed shells where the ``collective''
effects of pairing are expected to be reduced. Future experiments on $N=Z$ nuclei to 
determine the binding energies and the relative excitation energies
of $T=1$ and $T=0$ states (in odd-odd nuclei), as well as studies
of their high-spin rotational properties are necessary
and will provide further  tests of the role of pairing in $N=Z$ nuclei.

\vspace{0.5cm}
This work was supported under DOE contract No.~DE-AC03-76SF00098.
We are grateful to A.Goodman for discussions on the use of BCS equations that 
include np pairs.  One of us (AOM) would like to thank D.R.Bes, H.Sofia, N.Scoccola
and O.Civitarese for a valuable discussion on this subject.



\end{document}